# Origin of the Anisotropic Beer-Lambert Law from Dichroism and Birefringence in $\beta - Ga_2O_3$

Md Mohsinur Rahman Adnan[1], Mathias Schubert[3,4], and Roberto C. Myers[1,2*]

[1)] Department of Electrical and Computer Engineering, The Ohio State University, Columbus OH 43210, USA

[2)] Department of Materials Science Engineering, The Ohio State University, Columbus OH 43210, USA

[3)] Department of Electrical and Computer Engineering, University of Nebraska-Lincoln, Lincoln NE 68588, USA

[4)] NanoLund and Solid State Physics, Lund University, S-22100 Lund, Sweden

**The anisotropic optical absorption edge of β-Ga₂O₃ follows a modified Beer-Lambert law having two effective absorption coefficients. The absorption coefficient of linearly polarized light reduces to the least absorbing direction beyond a critical penetration depth, which itself depends on polarization and wavelength. To understand this behavior, a Stokes vector analysis is performed to track the polarization state as a function of depth. The weakening of the absorption coefficient is associated with a gradual shift of linear polarization to the least absorbing crystallographic direction in the plane, which is along the a-exciton within the (010) plane or along the b-exciton in the (001) plane. We show that strong linear dichroism near the optical absorption edge causes this shift in β-Ga₂O3, which arises from the anisotropy and spectral splitting of the physical absorbers i.e., excitons. The linear polarization shift is accompanied by a variation in the ellipticity due to the birefringence of β-Ga₂O₃. Analysis of the phase relationship between the incoming electric field to that at a certain depth reveals the phase speed as an effective refractive index, which varies along different crystallographic directions. The critical penetration depth is shown to be correlated with the depth at which the ellipticity is maximal. Thus, the anisotropic Beer-Lambert law arises from the interplay of both the dichroic and birefringent properties of β-Ga₂O₃.**

A fundamental assumption in the field of semiconductor optoelectronics is the applicability of the Beer-Lambert (BL) law, which states that while traversing an absorbing medium the photon flux ($\Phi$) decays exponentially with penetration depth ($z$), $\Phi = \Phi_0 e^{-\alpha z}$, where $\Phi_0$ is the photon flux at z = 0, and the exponent $\alpha$ is the medium's absorption coefficient [1]. However, this form of the BL law is only applicable for materials with plane wave Electro-Magnetic (EM) eigenmodes whose polarization states do not alter upon propagation to a certain distance. For highly anisotropic materials such as monoclinic $\beta - Ga_2O_3$, the eigenmode whose polarization state is unchanged upon propagation to a certain distance does not generally exist due to low symmetry. As a result, a strong deviation from the isotropic BL law is observed in such materials. In $\beta - Ga_2O_3$ the isotropic BL law has to be rewritten in a new formulation by incorporating dependence on the polarization angle ($\theta$), penetration depth ($z$) and energy ($E_{ph}$) of the photon. The new formulation of the BL law which explains the photon flux decay in this material can be written as [2],

$$\frac{\Phi}{\Phi_0} = \begin{cases} e^{-\alpha z}, & 0 < z \leq z_c \\ e^{-\alpha z_c} e^{-\alpha^*(z-z_c)}, & z_c < z \end{cases}, \tag{1}$$

Here, the flux decay follows the isotropic BL law with an absorption coefficient $\alpha$ until a critical penetration depth $z_c$ is reached. Both $\alpha$ and $z_c$ depend on the incoming wave's polarization angle $\theta$ and photon energy $E_{ph}$. Beyond $z_c$, the absorption coefficient $\alpha$ weakens to the smallest value of absorption coefficient $\alpha^*$ that is possible in the plane containing the EM fields. This is the anisotropic BL law which is observed both in a-c (010) and a-b (001) planes of $\beta - Ga_2O_3$ for the energy ranges $4.65 - 5.2\ eV$ and $4.9 - 5.8\ eV$ respectively. To understand the origin of this deviation from the isotropic BL law, the alteration of the polarization state of incoming wave upon traversing a certain distance ($z > z_c$) needs to be studied.

A monochromatic EM wave can be specified by the propagation vector $k$, the phase $\Phi$, and the polarization state, which is the curve traced out by the electric field $\vec{E}$ as a function of time in a fixed plane. The most general description of a fully, partially, or non- coherent polarization state is given by four parameters that describe the intensity (I), degree of polarization (p) and the shape parameters $(\psi, \chi)$ of the polarization ellipse. The Stokes vector is a set of values that describe the polarization state of an EM wave [3,4]. The relationship of the Stokes vector values to polarization state parameters are given as follows [5,6], $S_0 = I, S_1 = Ip\ cos2\psi\ cos2\chi, S_2 = Ip\ sin2\psi\ cos2\chi, S_3 = Ip\ sin2\chi$. Given the Stokes vector values, one can solve for the polarization state parameters as, $I = S_0, p = \frac{\sqrt{S_1^2+S_2^2+S_3^2}}{S_0}, 2\psi = \tan^{-1}\frac{S_2}{S_1}, 2\chi = \tan^{-1}\frac{S_3}{\sqrt{S_1^2+S_2^2}}$. Linear polarization preference direction parameter of the polarization ellipse is controlled by linear polarization shape parameter $\psi$, while elliptical shape parameter $\chi$ determines the perpetual change of ellipticity (linear to elliptical to circular to elliptical to linear) of the polarization state of the EM wave. In a fixed Cartesian $(\hat{x}, \hat{y})$ basis, the Stokes vector values are given as [5,6], $S_0 = |E_x|^2 + |E_y|^2, S_1 = |E_x|^2 - |E_y|^2, S_2 = 2\ Re(E_x E_y^*), S_3 = -2\ Im(E_x E_y^*)$ given $E_x$ and $E_y$ are the complex components of the electric field, i.e., $\vec{E} = \hat{x}E_x + \hat{y}E_y$. If the evolution of the $\vec{E}$ field as the wave propagates into a medium is known, then the polarization state at any depth can be determined by tracking the changes in the Stokes vector. In the case that the incoming light is linearly polarized along the x direction, i.e., $|E_x| = 1, |E_y| = 0$, the initial value of the linear polarization shape parameter is the same as the incoming wave's polarization angle $\psi = \theta$ and the initial value of elliptical shape parameter is $\chi = 0^0$.

Figure 1 shows the changes in the polarization state parameters $\psi$ in 1(a)-1(d) and $\chi$ in 1(e)-1(h) for light incident on the a-b (001) plane with energy range $E_{ph} = 4.9 - 5.8\ eV$ in 0.1 eV steps and crystal directions $\theta = 0^0, 30^0, 60^0$ and $90^0$. The a-axis is along $\theta = 0^0$ and the b-axis is along $\theta = 90^0$ while the other two angles represent non-major axes directions. For all these directions, $\psi$ and $\chi$ are plotted in $dz = 10\ nm$ steps from the surface ($z = 0$) to a depth of $z = 3000\ nm$. For the major axes a- and b- directions, the light polarization state does not change. As a result, $\psi = \theta = 0^0$ for the a-axis and $\psi = \theta = 90^0$ for the b-axis is always maintained. For both the major axes $\chi$ remains at $0°$ for all propagation distances. On the other hand, for polarizations along non-major axes directions, the linear polarization shape parameter is shifted towards $\psi = 90°$ which is equivalent to the b-axis direction $\theta = 90°$. The change is gradual and takes less steps to complete for $\theta = 60°$ compared to $\theta = 30°$. The b-axis in the a-b (001) plane is the dipole moment direction of the b-exciton ($X_b$) [7,8], thus light in this energy range becomes aligned to the dipole moment of $X_b$ after traversing a length of ~3 $\mu m$ in the material. This is true for all other linear polarizations that are not aligned with the major axes in a-b (001) plane. Phase variation, i.e., a gradual introduction of a phase shift a.k.a. ellipticity, is introduced just within the surface ($z > 0$) and the elliptical shape parameter $\chi$ starts to move away from $0°$ towards either being more positive or negative for both the non-major axes directions. After a range of $z$~3000 $nm$ the elliptical shape parameter returns to $0°$. This means that the ellipticity variation introduced just under the surface is not permanent and beyond a certain depth ~3 $\mu m$ the light will again be fully linearly polarized along the b-axis.

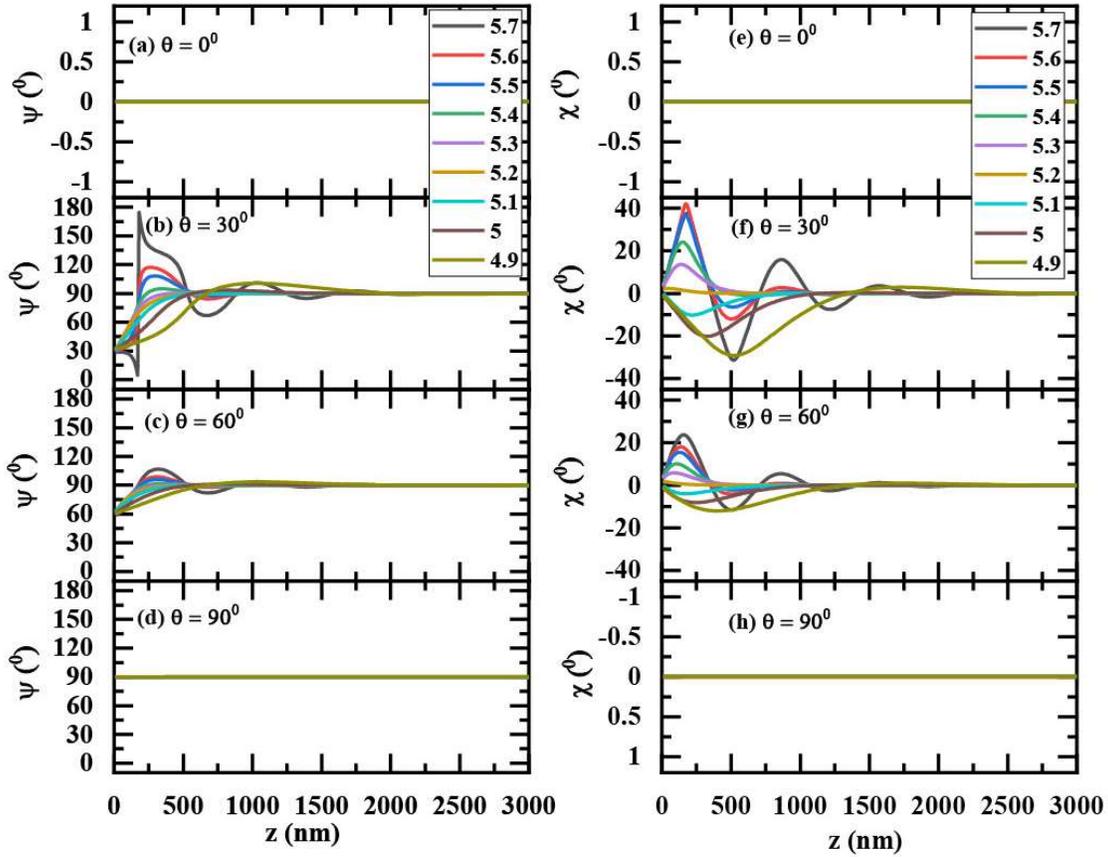

Figure 1: (a-d) Linear polarization shape parameter $\psi$ and (e-h) ellipticity shape parameter $\chi$ for light polarized within the a-b (001) plane with initial polarization direction $\theta$, incident on $\beta -$ $Ga_2O_3$ and propagating to a depth z for different photon energies (shown in the legends in units of eV). The polarization state for light along $\theta = 0°$ a- and $\theta = 90°$ b- axes does not vary with z, however, there is a gradual shift of $\psi$ towards the b-axis for other polarization states as well as ellipticity introduced near the surface, which decays with depth.

Figure 2 shows the changes in the polarization state parameters $\psi$ in 2(a)-2(d) and $\chi$ in 2(e)-2(h) for light incident on the a-c (010) plane with energy range $E_{ph} = 4.65 - 5.2\ eV$ in $0.1\ eV$ steps and polarization directions $\theta = 0°, 25.2°, 103.8°$ and $115.1°$. The a-axis is along $\theta = 0°$ and the c-axis is along $\theta = 103.8°$ while the other two angles represent the dipole moment directions of excitons in the a-c (010) plane. The $\theta = 25.2°$ light is polarized along the moment of a-exciton ($X_a$) while $\theta = 115.1°$ light is polarized along the moment of c-exciton ($X_c$) [7,8]. For all these directions, $\psi$ and $\chi$ are plotted in $dz = 10\ nm$ steps from the surface ($z = 0$) of $\beta - Ga_2O_3$ to a depth of $z = 10\ \mu m$. Unlike the a-b (001) plane, all directions including the major axes in a-c (010) plane experience both variations in the linear and elliptical shape parameters to the incoming polarization state. For all incoming polarization states the linear shape parameter variation takes place and a final polarization state is reached around $z \sim 10\ \mu m$. This final polarization state is identified to be along $\psi = \theta \cong 20° - 25°$ and has an energy dependence to it. For $4.9\ eV \leq E_{ph} < 5.2\ eV$, $\psi \sim 25°$ but in the range $E_{ph} < 4.9\ eV$, $\psi$ approaches a smaller value $\sim 20°$. Just like the a-b (001) plane, for all directions in the a-c (010) plane the ellipticity variation is introduced just within the surface ($z > 0$) and $\chi$ starts to move away from $0°$ for all directions. But the ellipticity variation parameter never settles down back to $\chi = 0°$, rather has some residual value even at z= $10000\ nm$ for all directions. So, even with fully linearly polarized light incident on $\beta - Ga_2O_3$ surface, a completely linear polarization state is never recovered within a depth of $\sim 10\ \mu m$ when the electric field of the travelling wave is oscillating in the a-c (010) plane. Nevertheless, for all directions, the $\chi$ remains very close to $0°$ (within $\pm 1°$) for $4.8\ eV \leq E_{ph} \leq 5\ eV$ when light has travelled deep enough $z \geq \sim 10\ \mu m$. Thus, an effectively linear polarized light along the dipole moment of $X_a$ is achieved in the a-c (010) plane at a depth of $\sim 10\ \mu m$ for this energy range.

In the anisotropic BL law for $\beta - Ga_2O_3$, the value of $\alpha^*$ was given as $\alpha^* = \alpha(\theta = 90°)$ in a-b (001) plane and $\alpha^* = \alpha(\theta \cong 25°)$ in a-c (010) plane. It is now established that the incoming wave's linear polarization state is shifted towards a specific direction in each plane of $\beta - Ga_2O_3$. These directions are oriented along $\theta = 90°$ in a-b (001) and $\theta \cong 20° - 25°$ in a-c (010) plane. So, the values of $\alpha^*$ matches $\alpha$ along these directions in the anisotropic BL law. Moreover, these directions are associated with exciton dipole moments in $\beta - Ga_2O_3$ which are oriented along $X_b$ at $90°$ from a-axis in a-b (001) and $X_a$ at a range of $\sim 17° - 25.2°$ from a-axis on a-c (010) plane. The third exciton $X_c$ is oriented at an angle in the range $\sim 110° - 115.1°$ from a-axis on a-c (010) plane [7,8]. These excitonic transitions are the three lowest near bandgap CP transitions including excitonic contributions [7,8] that take place at energies $X_c \sim 4.9\ eV, X_a \sim 5.2\ eV$ and $X_b \sim 5.5\ eV$. Since these transitions dominate the near bandgap photon absorption process [9,10] in $\beta - Ga_2O_3$, it is expected from the transition energies that in a-b (001) plane the least absorbing direction will be along $X_b$ ($\theta = 90°$) in the energy range $4.9 - 5.8\ eV$ and in a-c (001) plane the least absorbing direction will be along $X_a$ ($\theta \cong 20° - 25°$) in the energy range $4.65 - 5.2\ eV$. Thus, the physical basis of weakening of the absorption coefficient to the smallest value possible in plane is established to be a result of the anisotropy and energy shift of the physical absorbers, i.e., excitons.

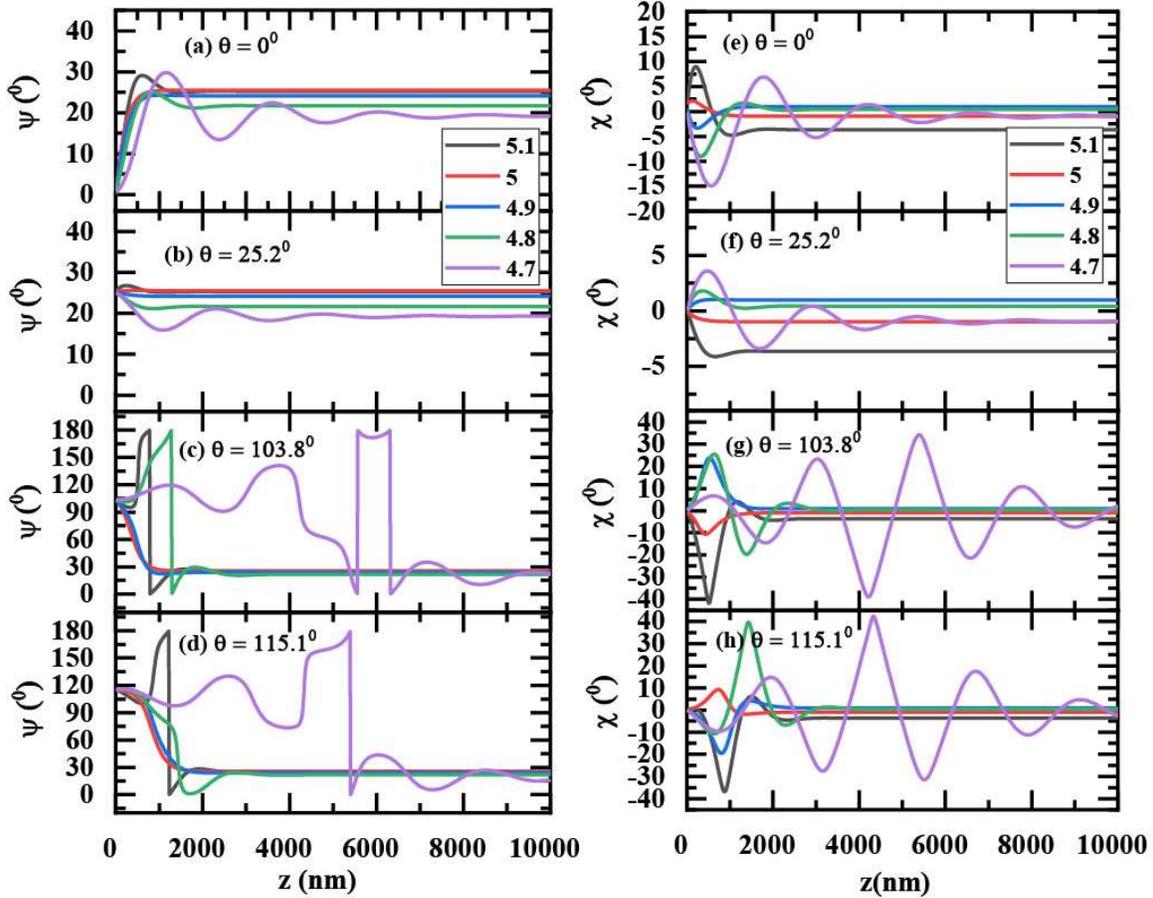

*Figure 2: (a-d) Linear polarization shape parameter ψ and (e-h) ellipticity shape parameter χ for light polarized within the a-c (010) plane with initial polarization direction θ, incident on $\beta - Ga_2O_3$ and propagating to a depth z for different photon energies (shown in the legends in units of eV). All polarization directions show a change in the polarization direction and develop ellipticity. The final linear polarization state shifts toward the direction if the dipole moment of the $X_a$ exciton at ψ ~20⁰ − 25⁰. The ellipticity persists to larger depths in the a-c (010) plane than in the a-b (001) plane (Fig. 1) before settling near 0⁰ for most of the energies in the range of interest.*

The gradual shift of ψ happens due to the dichroism property [2] of $\beta - Ga_2O_3$. Dichroism is the change in absorption of light with change in polarization direction [11]. Linear Dichroism ($L_D$) is defined as the difference in absorbance of light which are linearly polarized along orthogonal directions in a plane. Light aligned along any other polarization direction incident on the plane can be split into two components along these two orthogonal directions. Since one direction absorbs more than the other direction, the light component aligned along this more- absorbing direction will lose its amplitude faster than the other component. As a result, the sum of the components will gradually shift towards the less absorbing linear polarization direction.

A good illustration of this behavior is for linearly polarized light incident on the a-b (001) plane in $\beta - Ga_2O_3$. The two orthogonal directions are a- and b-axes and the $L_D$ in a-b (001) plane is shown in figure 5 of reference [2]. Since non-major axes polarized light's a-axis component is absorbed more [2] it will lose its amplitude faster than the component aligned along b-axis thus effectively shifting the non-major axes polarized light to b-axis after propagating some distance. Similarly, in a-c (010) plane since the light is absorbed more along the $X_c$ direction [2], the non-dipole-moment-orientation polarized light will eventually shift to the $X_a$ direction which is somewhat orthogonal to the $X_c$ direction [2]. The L$_D$ in a-b (001) plane has higher magnitude at any given depth compared to the a-c (010) plane. So, the change of $\psi$ to its final polarization state is completed earlier in a-b (001) plane at a depth of ~3 $\mu m$ compared to a-c (010) plane at a depth of ~10 $\mu m$.

The linear shape parameter shifting is accompanied by a strong variation of the ellipticity parameter. The depth required for the ellipticity to be removed and to recover a completely linear polarization state is larger for a-c (010) plane compared to a-b (001) plane. Even then, for a-c (010) plane the ellipticity is not completely removed rather has some residual value. The origin of this ellipticity parameter variation lies in the birefringence property [12,13] of $\beta - Ga_2O_3$. Birefringence is defined as the change in refractive index with change in polarization and propagation direction [14]. Refractive index, $n$ of a material is directly related to the velocity of light in the material. The refractive index is the ratio of light's velocity in vacuum, $c$ to that in the material, $v$ i.e., $n = \frac{c}{v} = \frac{ck}{\omega}$ where, $\omega$ is the angular frequency and $k$ is the propagation vector of the wave. Linear Birefringence (L$_B$) is the difference in refractive index of linearly polarized light along orthogonal directions in a plane. Then the L$_B$ between two orthogonal directions 1 and 2 will be defined as, $\Delta n = n_2 - n_1 = \frac{c(k_2-k_1)}{\omega}$. This means L$_B$ produces a relative difference in the propagation vector of the orthogonal components of the wave while passing through the plane. By doing this, in one direction light will move faster than the other and as a result there will be a retardation in phase angle between the orthogonal components aligned along these directions. Thus, a strong variation of the ellipticity parameter will be introduced.

To calculate the L$_B$, we need to find the refractive index along different directions in $\beta - Ga_2O_3$. The refractive index of a thin layer can be obtained by examining the phase relationship between the incoming and outgoing electric field for a given polarization utilizing the fact that the amount of phase retardation is related to the refractive index in the material. If the electric field for a linearly polarized light is given as, $|E| = E_0 e^{-i(\omega t - kz)}$, then the phase of the field is, $\Phi = \omega t - kz$. Just under the surface, any change in the electric field will be instantaneous, i.e., $t \sim 0$. If the wave propagates only a small distance, $\Delta z \sim 0$ within this $t \sim 0$ time then the change in phase will be, $\Delta \Phi = \Phi_{out} - \Phi_{in} = k(\Delta z - 0) = k\Delta z$. From this equation the propagation vector can be obtained as, $k = \frac{\Delta \Phi}{\Delta z}$. We know the velocity of light in a material is given as, $v = \frac{\omega}{k} = \frac{\frac{E_{ph}}{\hbar}}{\frac{\Delta \Phi}{\Delta z}} = \frac{E_{ph}\Delta z}{\hbar \Delta \Phi}$.

An effective refractive index can be obtained as, $n^* = \frac{c}{v} = \frac{c\hbar \Delta \Phi}{E_{ph} \Delta z}$ where $c$ is velocity of light in vacuum and $\hbar$ is Planck's constant. Thus, considering propagation perpendicular to the surface only, the phase speed can be obtained as an effective refractive index when the amount of phase retardation of the electric field after propagating a small distance into the material is known.

Figure 3 shows the effective refractive index ($n^*$) along different directions in both planes i.e., for the a-b (001) plane in 3(a)-3(c) and for the a-c plane (010) in 3(d)-3(f) at different small distances $\Delta z = 10\ nm, 20\ nm, 30\ nm$ respectively. Just below the bandgap [7,15] $E_g \sim 5.04\ eV$, $n^*$ is higher for the a-axis compared to the b-axis, while above bandgap $n^*$ is higher for the b-axis compared to the a-axis. Thus, the fast axis in the a-b (001) plane is energy dependent; below the bandgap it is the b-axis while above the bandgap it is the a-axis. Similarly, in the a-c (010) plane below the bandgap the a-axis is faster than the c-axis while above bandgap the c-axis is faster than the a-axis. Thus, just below the bandgap the order of the effective refractive index is $n_c^* > n_a^* > n_b^*$ but above the bandgap it is $n_b^* > n_a^* > n_c^*$. In the Near Infra-Red (NIR) range far below the bandgap the a-axis has the smallest refractive index, i.e., the highest light velocity in both planes, which is consistent with previous studies [12,16,17].

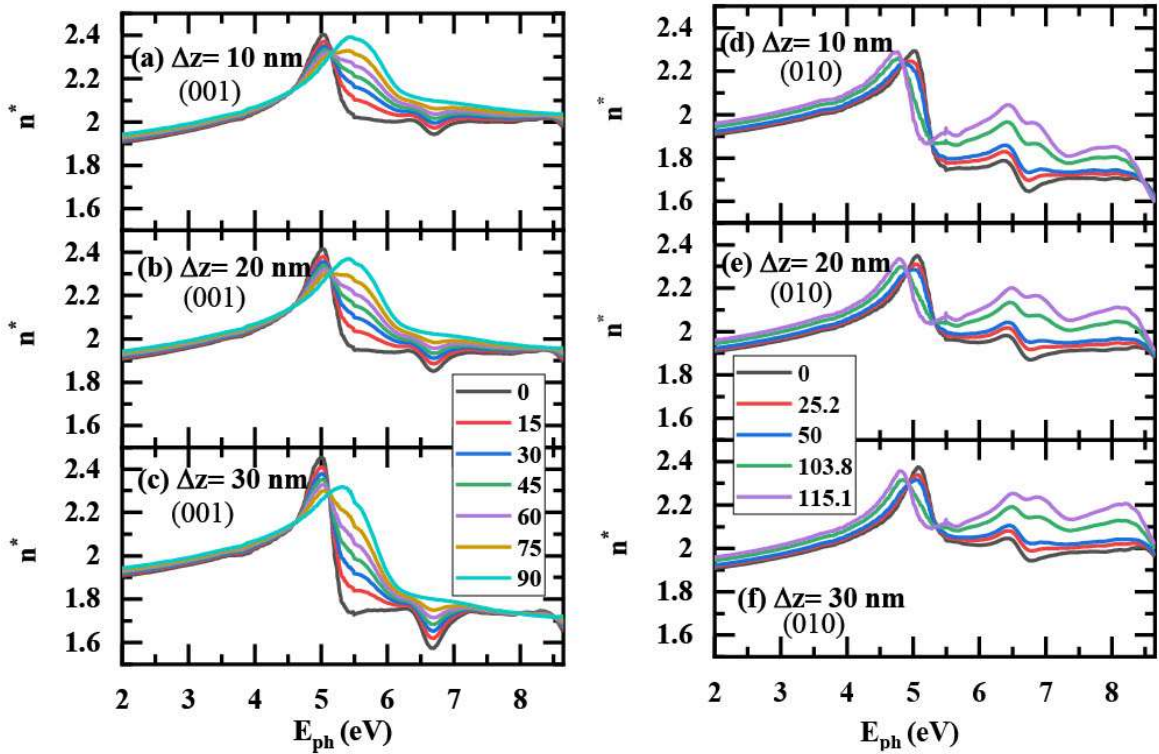

*Figure 3: Photon energy $E_{ph}$ dependence of the effective refractive index ($n^*$) for linearly polarized light incident on $\beta - Ga_2O_3$ at an angle $\theta$ (shown in the legends in units of degrees) within the (001) plane in parts (a-c), or within the (010) plane in parts (d-f), after propagating a total depth of $\Delta z$. The $n^*$ spectra are peaked near the bandgap for both planes. In the NIR range, the a-axis is always the fast axis regardless of the plane studied. In the vicinity of the excitonic transition energies the $n^*$ peaks broaden for the a-b (001) plane with increasing depth, while it narrows for the a-c (010) plane.*

The near surface $L_B$ is much larger in the a-b (001) plane than in the a-c (010) plane at any given distance $\Delta z \leq 30\ nm$, as shown in Figure 4, where the variation in $L_B$ for the a-b (001) plane is defined as $\Delta n^* = n_a^* - n_b^*$ and as $\Delta n^* = n_{X_c}^* - n_{X_a}^*$ for the a-c (010) plane. In the energy range of interest, the $L_B$ is seen to be moving towards zero with increasing $\Delta z$ for both planes. The maximum $L_B$ appears closest to the surface and reduces as light travels deeper. So, at larger depths $\beta - Ga_2O_3$ will act like an ideal linear polarizer with $L_D$ effects only. This explains why the elliptical shape of the polarization state is introduced just below the surface but vanishes at few microns depth. Since near the surface $L_B$ is much higher, the initial change in $\chi$ is much larger in the a-b (001) plane compared to the a-c (010) plane. The rate of change in $L_B$ is much higher in the a-b (001) plane compared to the a-c (010) plane. This is the reason of recovering zero ellipticity at a much smaller depth in the a-b (001) plane compared to that in the a-c (010) plane.

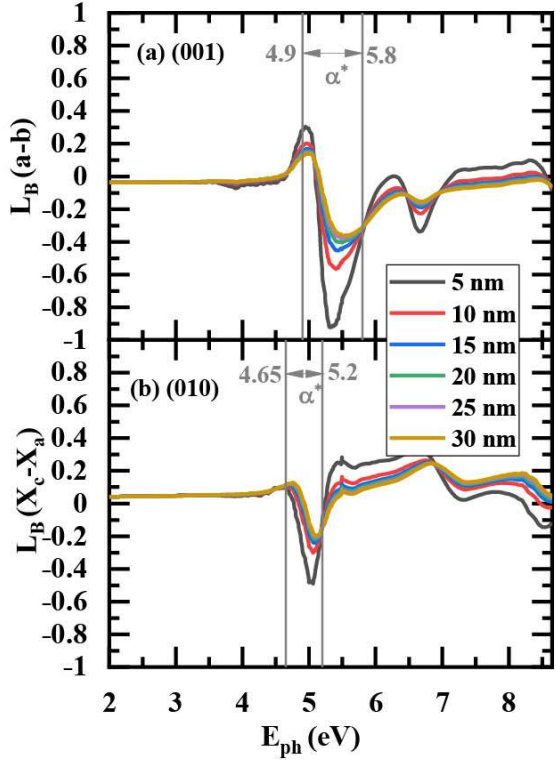

*Figure 4: (a,b) Photon energy $E_{ph}$ dependence of the linear birefringence $L_B$ for linearly polarized light incident on $\beta - Ga_2O_3$ within the (001) and the (010) plane, respectively, propagating a total depth of $\Delta z$ (shown in the legend). $L_B$ decays to 0 with increasing depth. The decay rate is larger for light polarized within the (001) plane compared with the (010) plane.*

The ellipticity parameter variation maximizes at different depths for different incoming polarization state and photon energy. The critical penetration depth $z_c$ has been found to be polarization angle and energy dependent [2]. Figure 5(a) shows the maximum of $\chi$ angle depth $z_m(\chi)$ as a function of energy for different polarization directions (dots) in the a-b (001) plane along with the critical penetration depth $z_c$ in the same energy range and same directions (line). Figure 5(b) shows the same results in the a-c (010) plane the same way. The critical penetration depth matches with maximum of $\chi$ angle depth in both planes for energy range $4.9\ eV \leq E_{ph} \leq 5.2\ eV$. For the a-b (001) plane the results match even in range $5.25\ eV \leq E_{ph} \leq 5.5\ eV$. Thus, for the energy range of the excitonic transitions, the maximum of $\chi$ angle depth $z_m(\chi)$ gives the critical penetration depth $z_c$. The maximum of $\chi$ angle is reached halfway between the initial and final linear polarization state, so this is the ideal depth for the least absorbing direction's absorption coefficient $\alpha^*$ to take over the incoming polarization state's absorption coefficient $\alpha$ in the anisotropic BL law.

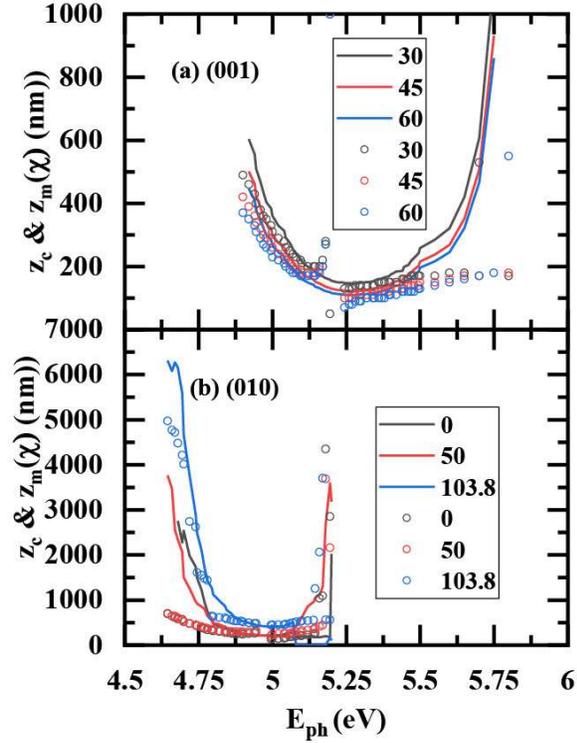

*Figure 5: (a,b) Photon energy $E_{ph}$ dependence of the maximum ellipticity depth $z_m(\chi)$ (open symbols) compared with critical penetration depth $z_c$ (lines) for linearly polarized light incident on $\beta - Ga_2O_3$ at an angle $\theta$ (shown in the legends in units of degrees) within the (001) plane and the (010) plane, respectively. The $z_m(\chi)$ values are determined from the Stokes vector analysis results shown in Figs. 1 and 2, whereas the $z_c$ values are based on electromagnetic modeling of the Poynting vector decay, i.e. fits to the anisotropic BL law described in Ref. [2].*

The anisotropic BL law arises from the dichroism and birefringence properties of $\beta - Ga_2O_3$. While the dichroism property induces a change in the linear polarization direction; the birefringence property induces ellipticity. The linear polarization shift dictates the final linear polarization state, which, in turn, dictates the value of the weakened absorption coefficient in the anisotropic BL law. The critical penetration depth reflects the balancing point beyond which the weaker absorption coefficient takes over, and coincides with the maximal ellipticity. We also obtain the energy and polarization dependence of the effective refractive index which provides the slow and fast axes over the wavelength range in $\beta - Ga_2O_3$ optoelectronic applications. Finally, the physical basis of the anisotropic absorption behavior in $\beta - Ga_2O_3$ is established to be originating from the anisotropy of the excitonic transitions. Such phenomena might have been overlooked and need thorough investigation in other materials with anisotropic excitons such as transition metal dichalcogenides and the entire class of low symmetry semiconductors.

*Conflict of Interest Statement- The authors have no conflicts to disclose.*

*Author Contributions- MMR Adnan performed numerical analysis, wrote the manuscript. M Schubert provided important insights about the obtained results, edited the manuscript. R Myers planned the study, edited the manuscript.*

*Data Availability Statement- The data that support the findings of this study are available from the corresponding author upon reasonable request.*

*Acknowledgements - This research was partially supported by the Center for Emergent Materials, an NSF MRSEC, under award number DMR-2011876. R.C.M acknowledges support by the Air Force*

*Office of Scientific Research (AFOSR) under awards FA9550-21-1-0278, FA9550-23-1-0330, and FA9550-24-1-0049. M.S. acknowledges support by the NSF under awards ECCS 2329940, and OIA-2044049 EQUATE, by AFOSR under awards FA9550-19-S-0003, FA9550-21-1-0259, and FA9550-23-1-0574 DEF, and by the University of Nebraska Foundation. M.S. also acknowledge support from the J.~A.~Woollam Foundation.*